\documentclass[12pt]{article}
\usepackage{graphicx}
\include{dspace12}
\begin{document}

\begin{flushright}
hep-ph/0605109 \\
HRI-P-06-05-002
\end{flushright}

\begin{center}
{\Large\bf Bilarge neutrino mixing in R-parity violating supersymmetry: the role of
right-chiral neutrino superfields}\\[20mm]

\renewcommand{\thefootnote}{\fnsymbol{footnote}}
Biswarup Mukhopadhyaya\footnote[1]{E-mail:
biswarup@mri.ernet.in}
and Raghavendra Srikanth\footnote[2]{E-mail:
srikanth@mri.ernet.in}\\

{\em Harish-Chandra
Research Institute,\\
Chhatnag Road, Jhusi, Allahabad - 211 019, India}

\end{center}

\begin{abstract}
We consider the possibility of neutrino mass generation in a
supersymmetric model where lepton number can be violated by 
odd units. The different patterns of mixing in the
quark and lepton sectors are attributed to the persence of
right-chiral neutrino superfields which (a) enter into Yukawa
couplings via non-renormalizable interaction with hidden sector
fields, and (b) can violate lepton number by odd units.
Both of these features are shown to be the result of some global
quantum number which is violated when SUSY is broken in the hidden
sector. It is shown how such a scenario, together with all known
R-parity violating effects, can lead to neutrino masses and bilarge
mixing via seesaw as well as radiative mechanisms. Some sample
values of the various parameters involved, consistent with electroweak
symmetry breaking constraints, are presented as illustrations.

\end{abstract}

PACS indices: 12.60.Jv, 14.60.Lm, 14.60.Pq, 14.60.St
\newpage

\section{Introduction}
It is expected that the upcoming accelerators operating around 
the TeV scale
will reveal some new principles in the domain of elementary particles. 
A frequently explored possibility in this context is supersymmetry (SUSY) 
\cite{susy}.
A strong motivation for postulating this additional symmetry  
is that SUSY can stabilize the
observed scale of electroweak symmetry breaking (EWSB) if it is broken
at or below the TeV region. The masses of the new particles in a
SUSY spectrum are thus expected to lie in this scale (commonly called
the `SUSY breaking scale'), although, in order
to achieve a consistent scheme, the origin of SUSY breaking is often
envisioned to lie in a higher energy range and a `hidden' sector.

There is no evidence of SUSY or any other kind of physics beyond the
standard model in collider experiments so far. The only strong hints
of `new physics', however, have been found at much lower energy scales,
in the world of neutrinos. If neutrino oscillations are indeed explanations
of the solar and atmospheric neutrino puzzles \cite{osci}, 
then one is bound to have
neutrino masses and mixing, which either require the presence of
right-handed neutrinos or necessitate lepton-number violation (or both,
as embodied in the seesaw mechanism). Furthermore, the mixing required
in the neutrino (or, more precisely,  leptonic) sector is of the 
bilarge type, with two large and one small mixing angles \cite{bilarge}.
This is quite different from quark mixing where one notices
progressively smaller mixing as one proceeds from the
first family to the second and the third. Understandably, this
makes one feel that some kind of physics beyond the standard model
(various theoretical possibilities have been explored in this connection
\cite{anarchy,a4,bmv,king,models})
is responsible for this difference in the neutrino (or, more precisely,
in the leptonic) sector.

If SUSY is indeed our key to  TeV-scale physics, could it also
be responsible for the novel features seen in the neutrino sector?
This question has been explored in a number of ways in recent times.
The two frequently discussed sources of neutrino mass generation 
are the seesaw mechanism and radiative effects. For the former in particular,
the scale of new physics, suppressing the relevant dimension-five operators,
normally has to be at least $10^{14}$ GeV, if neutrino masses are to have the
requisite order of smallness. A natural question to ask in this context is:
can we provide explanations of neutrino masses and mixing from the 
closest new physics scale around, such as that of SUSY breaking? 

The above question has already been addressed from various standpoints 
\cite{bmv,king,models,bm}.
In almost all of these approaches, it becomes necessary to postulate
some additional physics over and above the minimal SUSY standard model (MSSM).
However, the viability of a bilarge mixing pattern is not studied
with sufficient care in many of the existing approaches. We 
try to address this point, using  scales and symmetries that are 
invoked for ensuring a consistent SUSY breaking mechanism at the TeV
scale in the observable sector. The additional feature in this approach is the
inclusion of a right-chiral neutrino superfield for every family, something
that is inevitably required for a scheme like the seesaw mechanism. Such
a postulate has been been utilized earlier, where nonrenormalizable
interactions involving the right-chiral neutrinos and hidden sector
fields are included \cite{ahmsw1,ahmsw2,bn,by,mrs}. 
As can be seen in reference \cite{ahmsw1}, it is possible
to explain the value of the Higgsino mass parameter $\mu$,
obtaining it as an artifact of the breakdown of a global symmetry (R-symmetry)
at a scale of about $10^{11}$ GeV. It is interesting that the same 
broken global symmetry can also generate neutrino masses. 
It was argued in reference \cite{mrs}, using such a scenario,  that the special
nature of neutrino mixing is due to some terms in the
high-scale SUSY breaking scheme, including the right-chiral neutrino 
superfield, for which there is no analogue in the quark sector.  
While the viability of reproducing neutrino masses
in the correct range, using the intermediate scale of SUSY breaking
in the hidden sector, was successfully employed in earlier works \cite{ahmsw1},
later studies took a more comprehensive approach \cite{ahmsw2}, 
including radiative
as well as seesaw masses, and pointing out consistent regions in the
parameter space of the model answering to the bilarge mixing pattern 
\cite{mrs}.
The relative likelihood of the different neutrino mass scenarios, namely,
normal hierarchy, inverted hierarchy and degenerate neutrinos, could also
be studied in this approach \cite{mrs}. 

While our earlier study included $\Delta L = 2$ terms,
$\Delta L =1$ effects 
(or lepton number violation by an odd number in general) were
left out somewhat artificially. In other words, R-parity violating
effects were neglected in such a study. However, if lepton
number is violated in nature, there is little reason {\it a priori} in the 
claim that it is violated by only even units and not odd ones. 
The possibilities that open up on inclusion of R-parity violation are 
investigated in this paper.    

R-parity, defined by $R=(-1)^{3B+L+2S}$, is a conserved quantum number
in a SUSY theory so long as neither baryon nor lepton number is violated by
an odd number. However, while the gauge and current structures of the standard
model do not favour B/L violation, the situation is somewhat
different in SUSY.  Most importantly, the violation of R-parity  
does not necessarily cause unacceptable
consequences such as fast proton decay, if {\em only one} of
B and L is violated, a feature that is possible in SUSY since 
baryon and lepton numbers are carried by scalars as well. An important
phenomenological consequence of R-parity violation is that the lightest
SUSY particle (LSP) is not stable anymore. It has been shown 
\cite{rm,mrv,dmr,gh,jf,bem,hdprv,lm,kfc,bitri,brr} that
R-parity violation through lepton number can also give rise to neutrino 
masses in more than one ways, both by tree-level (seesaw type) effects
and radiative ones. 

Based on what has been said above, we have adopted the following programme
here:

\begin{itemize}
\item Use terms in the (nonrenormalizable) effective Lagrangian 
involving the MSSM superfields (including right-chiral neutrinos) 
and hidden sector superfields, some of which
are responsible for SUSY breaking, but including the possibility of
lepton number violation by odd units. Restrict the terms thus allowed by
some high-scale quantum number (R-charge).

\item Trigger SUSY breaking signaled by vacuum expectation values (vev)
acquired by the scalar as well as auxiliary components of the hidden
sector fields (whereby R-charge is broken).  Thus obtain the low-energy 
superpotential with broken R-parity, and also soft SUSY breaking terms
in the scalar potential.

\item Note that non-zero vev's for the sneutrinos can be generated
by the R-parity violating term(s) in the superpotential,
presumably around the TeV scale for the right-chiral ones but with
much smaller values for the left-chiral ones.

\item Using the low-energy effective Lagrangian and sneutrino vev's,
obtain seesaw as well as radiative masses for the light neutrinos,
and generate the neutrino mass matrix.

\item Equate terms of this mass matrix with that required by bilarge
mixing, and obtain constraints on the model parameters, taking into
account the conditions coming from electroweak symmetry breaking.

\item Show that numerically viable solutions exist, thus validating
the very postulate that an L-violating SUSY effective theory with 
right-chiral neutrino superfields can give rise to the observed 
mixing pattern.
\end{itemize}

The subsequent sections record different steps of this programme. In section
2 we describe the salient features of the high-scale theory and the form
of the low-energy Lagrangian once SUSY is broken. Section 3 specifies the 
requirements of bilarge neutrino mixing and generates the mass matrix
answering to such mixing in our scenario. Some typical numerical results 
are presented in section 4. We summarize and conclude in section 5. 

\section{The overall scenario}

As has been mentioned in the introduction, the scenario postulated here
attempts an extension of a recent work \cite{mrs} where we explained 
bilarge neutrino mixing pattern starting from nonremormalizable interactions 
which are induced from high-scale physics. The main feature
of this extension is that we now include lepton 
number violation by odd number of units. In the earlier study we used a 
global symmetry (called R-symmetry), whose purpose 
was to solve the $\mu$-problem \cite{ahmsw1,mup} and make the 
right-handed neutrino mass 
to be of order 1 TeV. However, since such R-charge is not an observed
quantum number in low-energy physics, one can assign it to fields 
differently compared to
the choices in \cite{mrs}. By thus identifying an appropriate  set of 
R-charges for both visible and hidden 
sector chiral superfields, we have found terms which violate lepton number 
by three units. We present these R-charges and the superpotential which 
is induced by high-scale physics in the next subsection.

The important thing to note is that the different R-charge of the right-chiral
neutrino superfield $N$ with respect to the other quarks and leptons sets it 
apart in its couplings to the hidden scetor. Consequently, $N$ in general
couples to different hidden sector fields as compared to the other chiral 
superfields, and the forms of these couplings are also different. This not only
results in a different role of $N$ in the superpotential but also  introduces
all the difference in the neutrino sector, in terms of both masses and mixing.

\subsection{Superpotential}

\begin{table}
\begin{center}
\begin{tabular}{|c|c|cccccccc|}  \hline

Hidden sector & Field & $X_{ij}$ & $\bar{X}_{ij}$ &  
$S_{ij}$ & $Z_{ij}$ & $\bar{Z}_{ij}$ & & &
  \\ \cline{2-2}
 & R-charge & $-\frac{1}{6}$ & $\frac{1}{6}$ &  
2 & $\frac{5}{3}$ & $\frac{7}{3}$ & & & \\  \hline
Visible sector & Field & $Q_i$ & $L_i$ & $U^c_i$ & $D^c_i$ & $E^c_i$ 
& $N_i$ & $H_u$ & $H_d$     \\ \cline{2-2}
 & R-charge & $\frac{1}{2}$ & $\frac{1}{2}$ & $\frac{1}{2}$ & $\frac{1}{2}$ &
$\frac{1}{2}$ & $\frac{2}{3}$ & 1 & 1  \\ \hline

\end{tabular}
\end{center}
\caption{R-charges of hidden and visible sector superfields.}
\end{table}

The proposed superpotential for the chiral superfields in the observable sector
has the following form before SUSY breaking:
\begin{equation}
W = Y^{ij}_uQ_iU^c_jH_u + Y^{ij}_dQ_iD^c_jH_d + Y^{ij}_eL_iE^c_jH_d + 
\frac{1}{M_P}X^{ij}L_iN_jH_u + \kappa^{ijk}N_iN_jN_k.
\end{equation}
where $N_i$ (i = 1-3) correspond to the three right-chiral neutrino 
superfields, and the $X_{ij}$ constitute an array of hidden sector chiral 
supefields which can couple to the chiral superfields only in terms
where the $N_i$ fields are involved. This happens by virtue of the conserved 
global quantum number R. 
$\mu$-term in the superpotential is generated through fields like 
$S_{ij}$ and that is explained in \cite{mrs}. 

When SUSY is broken in the hidden sector, F-terms of certain fields 
are in general responsible. However, as we have already warned
the reader, the fields $X_{ij}$ have a slightly different stature, in the
sense that it has a special R-charge, so as to couple to observable
sector fields via non-renormalizable interactions only when
the $N_i$  are present in the interaction. Moreover,
non-zero vev's are acquired by 
the scalar ($\langle x_{ij}\rangle$) but {\it not} the  auxiliary 
($\langle {F_X}_{ij}\rangle$) components of $X_{ij}$ \footnote{Of course,
we require some non-vanishing F-terms to break SUSY in the hidden sector.
Such F-terms are attributed to other hidden sector fields, who have
the right R-charges to give masses to the usual squarks, sleptons and 
gauginos.}. R-charge is also broken at this scale.
The fourth term of the above superpotential generates Yukawa couplings 
(and subsequently Dirac mass terms when electroweak symmetry is broken) 
for neutrinos.

The process of acquiring vev's requires  $X_{ij}$ to be coupled with
other hidden sector superfields. In this model, such fields are taken
to be the arrays  $Z_{ij}$ and   $S_{ij}$ (taking
the hidden sector fields to be symmetric in $i,j$, in order to keep
the analysis simple). 
The R-charges required by these as well as the observable sector
superfields for overall consistency 
are listed in table 1. In a similar way as in reference 
\cite{ahmsw1,ahmsw2,mrs}, 
the relevant part of the superpotential containing these fields can be 
expressed as

\begin{equation}
{\cal W} = \sum_{i,j=1}^{3} S_{ij} (X_{ij}\bar{X}_{ij} - \mu^{2}_{ij}) +
         X_{ij}^2\bar{Z}_{ij} + \bar{X}_{ij}^2Z_{ij}.
\end{equation}
The scalar potential arising out of the above superpotential, after 
minimization, gives vev's to the scalar and auxillary components of the
superfields involved, thus breaking SUSY and R-charge. In particular,
the vev's acquired by the components of $X_{ij}$ modify the low-energy
Lagrangian in the observable sector, as evident from the superpotential
shown in equation (1).

The above procedure allows us to have 
$\langle F_{X_{ij}}\rangle = 0$ and $\langle x_{ij}\rangle \neq 0$
for all $i,j$. The vanishing off-dagonal $F_X$ vev's
ensure the suppression of flavour changing neutral currents (FCNC).
By making the diagonal components zero, one ensures the radiative
contributions to neutrino masses are not 
inadmissibly high (see section 3.3).

In addition, the low-energy observable sector superpotential requires
the inclusion of the term $\mu H_d H_u$. This term 
is generated via interactions of the form 
$\sum_{i,j}\int d^4\theta\frac{S^\dagger_{ij}}{M_P}H_dH_u$, which
are R-charge conserving. It leads to the the usual $\mu$-term with
$\mu \simeq$ TeV if $\sum_{i,j} \langle {F_S}_{ij}\rangle \simeq 10^{22} 
~{\rm GeV^2}$.

Thus, after SUSY and R-charge breaking in the hiddeen sector,
the observable sector superpotential reduces to the form 
\begin{equation}
W = Y^{ij}_uQ_iU^c_jH_u + Y^{ij}_dQ_iD^c_jH_d + Y^{ij}_eL_iE^c_jH_d +
Y^{ij}_{\nu}L_iN_jH_u - \mu H_dH_u + \kappa^{ijk}N_iN_jN_k,
\end{equation}

\noindent
where $Y^{ij}_{\nu} = \frac{\langle x_{ij}\rangle}{M_P}$, giving 
symmetric Yukawa couplings in the flavour space for neutrinos.  
This is basically the MSSM superpotential plus the Yukawa coupling term 
for the neutrinos and the $\Delta L = 3$ term which makes
it R-parity violating. It is this term \cite{lm}, trilinear in $N$, 
which provides
the crucial distinction of this model with the R-parity conserving case
presented in reference \cite{mrs}. This term provides the origin of 
$\Delta L = 2$
masses for both right-chiral neutrinos and sneutrinos, as opposed to the  
situation \cite{mrs}, and that is why zero F-terms as well as different 
R-charges have been assigned to the $X$-fields.

The above observations reveal a rather interesting feature of the model.
In the observable sector, we are convinced that neutrinos are somewhat 
different from other fermions, as revealed not only by their much smaller 
masses but also in their completely different mixing pattern. We attribute
this to the special nature of the right-chiral neutrino superfields \cite{gh},
possessing different R-charges compared to all other chiral superfields. 
Such a different R-charge enables them to couple to a different, special 
set of hidden sector fields, namely, the $X_{ij}$. It turns out that the 
$X_{ij}$ also have a special property, in the sense that their auxiliary 
components have zero vev's. This feature, and the fact that the $N_i$, 
carrying their L-violating terms into the superpotential, are coupled
with them, makes the neutrino sector quite distinct from the
remaining fermions. What is especially interesting is that all this can happen
with the right-handed neutrino masses as well as the vev's of the corresponding
sneutrinos in  the Tev-scale.

As we shall see in the next subsection, if we write down the scalar potential
from equation (3) and add the appropriate D-terms and soft SUSY-breaking terms,
then the  sneutrino fields, both left-and right-chiral, 
acquire vev's after minimization of the potential. 
Tree-level neutrino masses can be consequently generated from 
neutralino-neutrino mixing, and one can argue that if such masses 
have to be of the right order of magnitude, then the vev of the left-chiral 
sneutrinos should not exceed $10^{-4}$ GeV \cite{rm,dmr,jf,lm}. This is
because tree-level neutrino mass generation is a seesaw type effect. Unless
the left-chiral sneutrino vev is small enough, one cannot produce 
eigenvalues with the required degree of smallness unless one `aligns' the
$\mu$-parameter and the vector $\bar{\epsilon} = (\epsilon_1,\epsilon_2,
\epsilon_3)$, $\epsilon_i$ being the coefficient of the effective bilinear
term $L_iH_2$ in the superpotential \cite{sacha}. In the absence of a
symmetry postulated to ensure such an alignment, it therefore makes sense
to accept small values of $\langle\tilde{\nu}_i\rangle$ as a phenomenological
constraint.

The value of the right-chiral sneutrino vev ($\langle\tilde{n}_i\rangle$),
however, can be in the TeV scale, since $\langle\tilde{n}_i\rangle$ is
responsible for the right-handed Majorana masses. This is compatible with
small values of the effective $\epsilon_i$ if $Y_{\nu}\sim 10^{-7}$. Also,
it is possible to set all three $\langle\tilde{n}_i\rangle$ at the same
value without any loss of generality, as we have done later in this paper.

We simplify our 
algebra by rotating away these small vev's via a basis change. To make
the matter clear, let us assume that in the original basis the neutral 
scalar fields acquire non-zero vev's as follows:
\begin{eqnarray}
\langle H^0_u\rangle = v_u, \quad \langle H^0_d\rangle = v_1, \quad 
\langle \tilde{\nu}_i\rangle = \langle {\nu}_i\rangle ,\quad 
\langle \tilde{n}_i\rangle = n_i.
\end{eqnarray}
Define 
\begin{equation}
v_d = \sqrt{v_1^2 + \sum_i\langle {\nu}_i\rangle^2}.
\end{equation}
Once lepton number violation is allowed, we can switch to  
a new basis $H_d^\prime ,L_i^\prime$ through the transformation matrix $O$: 
\begin{equation}
\left(\begin{array}{c}
H_d \\
L_i 
\end{array}\right) = O
\left(\begin{array}{c}
H_d^\prime \\
L_i^\prime 
\end{array}\right),
\end{equation}
where 
\begin{equation}
O = \frac{1}{v_d}
\left(\begin{array}{cccc}
v_1 & -\langle \nu_1\rangle & -\langle \nu_2\rangle & -\langle \nu_3\rangle \\
\langle \nu_1\rangle & v_1 & \langle \nu_3\rangle & -\langle \nu_2\rangle \\
\langle \nu_2\rangle & -\langle \nu_3\rangle & v_1 & \langle \nu_1\rangle \\
\langle \nu_3\rangle & \langle \nu_2\rangle & -\langle \nu_1\rangle & v_1 
\end{array}\right).
\end{equation}
giving $\langle H^{0\prime}_d\rangle = v_d$, 
$\langle \tilde{\nu}_i\rangle = 0$ in the new basis. 
In this basis the superpotential becomes
\begin{eqnarray}
W &=& Y^{ij}_uQ_iU^c_jH_u + Y^{ij}_dO_{00}Q_iD^c_jH_d^\prime + 
Y^{ij}_dO_{0k}Q_iD^c_jL_k^\prime + 
	\nonumber \\
&& Y^{kj}_e(O_{ki}O_{00}+O_{k0}O_{oi})L_i^\prime E^c_jH_d^\prime +
(Y^{lk}_eO_{li}O_{0j})L_i^\prime L_j^\prime E^c_k + 
Y^{kj}_\nu O_{ki}L_i^\prime N_jH_u + 
	\nonumber \\
&& Y^{ij}_\nu O_{i0}H_d^\prime H_uN_j 
- \mu O_{00}H_dH_u^\prime - \mu O_{0i}L_i^\prime H_u + 
\kappa^{ijk}N_iN_jN_k,
\end{eqnarray}
where the indices in $O_{\alpha\beta}$ take the values (0,i) with i = 1-3.
The first index coresponds to $H_d^\prime$ and the remaining ones, to 
$L_i^\prime$. It is to be 
noticed that, once the phenomenological constraints are imposed,
$O_{0i},O_{i0}\sim\frac{\langle \nu_i\rangle}{v_d}\sim 10^{-6}$, 
$O_{00}\approx 1$ and $O_{ij}\approx\delta_{ij}$. 

The small values of $Y^{ij}_\nu$ imply that the terms with  
$H_d^\prime H_uN_j$ can be neglected. Those involving
$Q_iD^c_jL_k^\prime$ and $L_i^\prime L_j^\prime E^c_k$ are the usual trilinear
L(and therefore R-parity)-violating terms. 
Putting $\epsilon_i = \mu O_{0i}$, we also recover the
bilinear R-parity violating effects, where the information of sneutrino 
vev's before rotation has gone in. 

Now, if we drop the primes in $H_d^\prime ,L_i^\prime$, then  
the superpotential finally takes the form 
\begin{eqnarray}
W &=& Y^{ij}_uQ_iU^c_jH_u + Y^{ij}_dQ_iD^c_jH_d + Y^{ij}_eL_iE^c_jH_d +
Y^{ij}_{\nu}L_iN_jH_u + \lambda^{ijk}L_iL_jE^c_k + 
\lambda^{\prime ijk}Q_iD^c_jL_k
	\nonumber \\
&& - \mu H_dH_u - \epsilon_iL_iH_u + 
\kappa^{ijk}N_iN_jN_k,
\end{eqnarray}
where $\lambda^{ijk}=Y^{lk}_eO_{li}O_{0j}$ and 
$\lambda^{\prime ijk}=Y^{ij}_dO_{0k}$, which is strikingly close to the 
most general R-parity violating
superpotential usually found in the literature \cite{bgh}, now derived from
physics in the hidden sector. In addition, we have the terms trilinear
in the N-superfields. Two very important roles played by this term are
(a) developing non-zero vev's ($n_i$) for the right-chiral sneutrinos, and
(b) the generation of $\Delta L =2$ right-handed neutrino masses.
As will be demonstrated in the next section, $n_i$ is within the TeV scale
in a theory of this kind. The right-handed Majorana mass terms are then also
in the same scale, and thus our ambition of explaining neutrino physics
by the SUSY breaking scale is furthered by the scenario constructed here.

\subsection{The scalar potential and electroweak symmetry breaking  
conditions}

The electrically neutral part of the scalar potential, 
consisting of F-terms induced by the 
above superpotential, D-terms as well as soft SUSY breaking terms, is 

\begin{eqnarray}
V = |Y^{ij}_{\nu}\tilde{\nu}_i\tilde{N}_j - \epsilon_i\tilde{\nu}_i - 
\mu H_d^0|^2 + |\mu H_u^0|^2 + \sum_i|Y^{ij}_{\nu}H_u^0\tilde{N}_j - 
\epsilon_iH_u^0|^2 + \sum_i|Y^{ji}_{\nu}\tilde{\nu}_jH_u^0 + 
3\kappa^{ijk}\tilde{N}_j\tilde{N}_k|^2 + 
	\nonumber \\
(\frac{g^{\prime 2}+g^2}{8})
[|H_u^0|^2 - |H_d^0|^2 - \sum_i|\tilde{\nu}_i|^2]^2 + m_{H_d}^2|H_d^0|^2 + 
m_{H_u}^2|H_u^0|^2 + (m_{\tilde{L}}^2)^{ij}\tilde{\nu}_i^*\tilde{\nu}_j + 
(m_{\tilde{N}}^2)^{ij}\tilde{N}_i^*\tilde{N}_j + 
	\nonumber \\ 
\left[(A_\nu Y_\nu )^{ij}
\tilde{\nu}_i\tilde{N}_jH_u^0 - (B_\mu\mu) H_u^0H_d^0 - (B_\epsilon\epsilon )^i
\tilde{\nu}_iH_u^0 + \frac{1}{3}(A_\kappa \kappa )^{ijk}\tilde{N}_i
\tilde{N}_j\tilde{N}_k + {\rm H.c.}\right].
\end{eqnarray}

There are eight neutral scalar fields in our model, which can develop
non-zero vev's. Thus after EWSB we expect
\begin{eqnarray}
\langle H_u^0 \rangle = v_u, \quad \langle H_d^0 \rangle = v_d, \quad 
\langle \tilde{\nu}_i\rangle = 0, \quad 
\langle \tilde{N}_i\rangle = n_i.
\end{eqnarray}
In order to show that right-chiral sneutrino vev 
is on the order of a TeV, we have to study conditions that arise from
the minimization of the potential \cite{rm,dmr,ewsb}. The first thing to 
notice is that 
the potential is bounded from below because 
the fourth powers of all the eight neutral fields are positive. The extremal
conditions with respect to the various fields are 
\begin{eqnarray}
\mu^2v_d - \frac{g^{\prime 2}+g^2}{4}(v_u^2-v_d^2)v_d + m_{H_d}^2v_d -
(B_\mu\mu)v_u &=& 0,
	\nonumber \\
\mu^2v_u + \sum_i(Y^{ij}_\nu n_j-\epsilon_i)(Y^{ik}_\nu n_k-\epsilon_i)v_u + 
\frac{g^{\prime 2}+g^2}{4}(v_u^2-v_d^2)v_u + m_{H_u}^2v_u - 
(B_\mu\mu)v_d &=& 0,
	\nonumber \\
\sum_lY^{li}_\nu (Y^{lj}_\nu n_j-\epsilon_l)v_u^2 + \sum_l(6\kappa^{lik}n_k)
(3\kappa^{ljk}n_jn_k) + (m_{\tilde{N}})^{ji}n_j + (A_\kappa\kappa )^{ijk}
n_jn_k &=& 0,
	\nonumber \\
-\mu(Y^{ij}_\nu n_j-\epsilon_i)v_d + \sum_l(Y^{il}_\nu v_u)
(3\kappa^{lik}n_jn_k) + (A_\nu Y_\nu )^{ij}n_jv_u - (B_\epsilon\epsilon )^i
v_u &=& 0.
\end{eqnarray}
In deriving the above equations, we have assumed $\kappa^{ijk}$ and
$(A_\kappa\kappa )^{ijk}$ to be symmetric in $i,j,k$. 

Moreover, we need to ensure that the extremum value coresponds to  
the minimum of the potential, by studying the 
second derivatives, given as
\begin{eqnarray}
\frac{1}{2}\frac{\partial^2V}{\partial H_d^{0^2}} &=& 
\mu^2 - \frac{g^{\prime 2}+g^2}{4}(v_u^2-3v_d^2) + m_{H_d}^2,
%	\nonumber \\
\quad
\frac{1}{2}\frac{\partial^2V}{\partial H_d^0\partial H_u^0} ~=~
-\frac{g^{\prime 2}+g^2}{2}v_uv_d - (B_\mu\mu ),
	\nonumber \\
\frac{1}{2}\frac{\partial^2V}{\partial H_d^0\partial\tilde{N}_i} &=&
0,
%	\nonumber \\
\quad
\frac{1}{2}\frac{\partial^2V}{\partial H_d^0\partial\tilde{\nu}_i} ~=~
-\mu(Y^{ij}_\nu n_j - \epsilon_i),
	\nonumber \\
\frac{1}{2}\frac{\partial^2V}{\partial H_u^{0^2}} &=&
\mu^2 + \sum_i(Y^{ij}_\nu n_j - \epsilon_i)(Y^{ik}_\nu n_k - \epsilon_i) + 
\frac{g^{\prime 2}+g^2}{4}(3v_u^2-v_d^2) + m_{H_u}^2,
	\nonumber \\
\frac{1}{2}\frac{\partial^2V}{\partial H_u^0\partial\tilde{N}_i} &=&
2\sum_lY^{li}_\nu (Y^{lk}_\nu n_k - \epsilon_l)v_u,
%	\nonumber \\
\quad
\frac{1}{2}\frac{\partial^2V}{\partial H_u^0\partial\tilde{\nu}_i} ~=~
\sum_l3Y^{il}_\nu \kappa^{ljk}n_jn_k + (A_\nu Y_\nu )^{ij}n_j - 
(B_\epsilon\epsilon )^i,
	\nonumber \\
\frac{1}{2}\frac{\partial^2V}{\partial\tilde{N}_i\partial\tilde{N}_j} &=&
\sum_lY^{li}_\nu Y^{lj}_\nu v_u^2 + \sum_l(6\kappa^{lij}3\kappa^{lmk}n_mn_k 
+ 6\kappa^{lik}n_k6\kappa^{ljm}n_m) + (m_{\tilde{N}}^2)^{ji} + 
2(A_\kappa\kappa )^{ijk}n_k,
	\nonumber \\
\frac{1}{2}\frac{\partial^2V}{\partial\tilde{N}_i\partial\tilde{\nu}_j} &=&
-Y^{ji}_\nu \mu v_d + \sum_l6\kappa^{lik}Y^{jl}_\nu n_kv_u + 
(A_\nu Y_\nu )^{ji}v_u,
	\nonumber \\
\frac{1}{2}\frac{\partial^2V}{\partial\tilde{\nu}_i\partial\tilde{\nu}_j} &=&
(Y^{il}_\nu n_l - \epsilon_i)(Y^{jk}_\nu n_k - \epsilon_j) + 
\sum_lY^{il}_\nu Y^{jl}_\nu v_u^2 - \frac{g^{\prime 2}+g^2}{4}(v_u^2-v_d^2)
\delta_{ij} + (m_{\tilde{L}}^2)^{ji}.
\end{eqnarray}
The above set of equations give an 8$\times$8 symmetric mass-squared matrix. 
All the eight eigenvalues of this matrix should come as positive for
a minimum. 

The other condition that has been employed here is that the potential
should be bounded from below in the direction $H^0_u = H^0_d$, along which
the quadratic term must have a positive coefficient. This conditions gives
$2(\mu^2-(B_\mu\mu))+m_{H_d}^2+m_{H_u}^2 \geq 0$.
Finally, the potential evaluated at the minima 
should be less than zero.

All the above conditions have been imposed on the scalar potential
in order to constrain the various parameters here. We have first 
assumed that the soft SUSY breaking parameters are within a TeV or so.
Next, quanities which can potentially contribue to neutrino masses (such
as $\epsilon_i$ and the Yukawa couplings $Y^{ij}_\nu$) are subject to
additional constraints. Using all these, and the full set of minimization 
conditions, one finds that the right-chiral sneutrino vev's $n_i$ come 
out consistently in the TeV range.

\section{Neutrino masses}

\subsection{General features}

Experimentally three mass eigenstates of neutrinos have been found so far,
and, according to Z-decay results, there canot be any more light sequential
neutrinos. Thus the light neutrinos form a 3$\times$3 mass matrix in the 
flavour basis. The unitary matrix which diagonalizes this mass matrix can be
parameterized as \cite{pardata}
\begin{equation}
U = \left( \begin{array}{ccc}
        c_{12}c_{13} & s_{12}c_{13} & s_{13}e^{-i\delta } \\
        -s_{12}c_{23}-c_{12}s_{23}s_{13}e^{i\delta } &
        c_{12}c_{23}-s_{12}s_{23}s_{13}e^{i\delta } & s_{23}c_{13} \\
        s_{12}s_{23}-c_{12}c_{23}s_{13}e^{i\delta } &
        -c_{12}s_{23}-s_{12}c_{23}s_{13}e^{i\delta } & c_{23}c_{13}
     \end{array} \right),
\end{equation}
where $c_{ij} = \cos\theta_{ij}$, $s_{ij} = \sin\theta_{ij}$ and $i,j$ run 
from 1 to 3. Various neutrino 
oscillation experiments indicate that
$\theta_{12}\approx 32^{\rm o}$, $\theta_{23}\approx 45^{\rm o}$ and 
$\theta_{13}\leq 13^{\rm o}$ \cite{mixang,sv}. This pattern is known as 
bilarge mixing. 
In order to understand the consequences of such mixing in the zeroth order,
we can approximately take $\theta_{23} = \frac{\pi}{4}$, $\theta_{13} = 0$ 
and $\sin\theta_{12} = \frac{1}{\sqrt{3}}$, something known as tri-bimaximal 
structure \cite{hps}. Then the unitary matrix turns out to be 
\begin{equation}
U_\nu = \left( \begin{array}{ccc}
        \sqrt{\frac{2}{3}} & \frac{1}{\sqrt{3}} & 0 \\
        -\frac{1}{\sqrt{6}} & \frac{1}{\sqrt{3}} & \frac{1}{\sqrt{2}} \\
        \frac{1}{\sqrt{6}} & -\frac{1}{\sqrt{3}} & \frac{1}{\sqrt{2}}
        \end{array} \right).
\end{equation}
The effects of a small but non-zero $\theta_{13}$ is subleading and they 
do not change our conclusions qualitatively.
Given the three light mass eigenvalues $m_1$, $m_2$ and $m_3$, it is possible
to use the matrix $U_\nu$ to obtain the mass matrix in the flavour basis. For 
Majorana neutrinos, in particular, the mass matrix can be written as 
\begin{eqnarray}
m_\nu &=& U_\nu \left( \begin{array}{ccc}
        m_1 & & \\
         & m_2 & \\
         & & m_3
        \end{array} \right) U^T_\nu
                \nonumber \\
      &=& \left( \begin{array}{ccc}
        \frac{1}{3}(2m_1+m_2) & \frac{1}{3}(-m_1+m_2) &
        \frac{1}{3}(m_1-m_2) \\
        \frac{1}{3}(-m_1+m_2) & \frac{1}{6}(m_1+2m_2+3m_3) &
        \frac{1}{6}(-m_1-2m_2+3m_3) \\
        \frac{1}{3}(m_1-m_2) & \frac{1}{6}(-m_1-2m_2+3m_3) &
        \frac{1}{6}(m_1+2m_2+3m_3)
        \end{array} \right),
\end{eqnarray}

We shall next examine how a light neutrino mass matrix of the above
type can be generated in the scenario proposed here. For that, we take
up the cases of seesaw and radiative masses in the next two subsections.

\subsection{Seesaw masses}

When lepton number violation is allowed, the light neutrino mass matrix 
arising via the seesaw mechanism  is in general given by 
\begin{equation}
m_{\nu}^s = -mM^{-1}m^T,
\end{equation}
where $m$ is the so-called Dirac neutrino mass matrix, basically representing
the terms bilinear in the light and heavy degrees of freedom. $M$ is the
mass matrix for the heavy states. While $M$ consists of $\Delta L =2$ mass
terms for right-handed neutrinos in the usual seesaw mechanism, in our case 
$M$ contains the neutralino mass matrix as well. This is because our
superpotential admits neutrino-neutralino mixing \cite{rm,gh,lm}. 
Thus, with three
right-handed neutrinos and four neutralino states, $M$ is a $7\times 7$
matrix here. It should be noted that $M$ is block diagonal in the basis
where the left sneutrino vev's are rotated away.
$m$, on the other hand, is a $3\times 7$ matrix, including the
`real' Dirac mass terms $\bar{\nu}_L N_R$ as well as the neutrino-neutralino
mixing terms driven by the quantities $\epsilon_i$  in the same basis.

Thus, in the basis 
\begin{equation}
\psi_M = (\tilde{B}, \tilde{W}^3, \tilde{H}_d^0, \tilde{H}_u^0, N_i, \nu_i)
\end{equation}
we get mass terms of the form
\begin{equation}
{\cal L}_m = -\frac{1}{2}\psi_M^TM_n\psi_M + {\rm H.c.},
\end{equation}
where
\begin{equation}
M_n = 
\left(\begin{array}{cc}
M_{7\times 7} & m_{7\times 3}^T \\
m_{3\times 7} & 0
\end{array}\right).
\end{equation}
The TeV scale matrix formed by neutralinos and right-handed neutrinos is 
\begin{equation}
M_{7\times 7} = 
\left(\begin{array}{cc}
M_{\chi^0} & 0 \\
0 & M_{N-N} 
\end{array}\right),
\end{equation}
where
\begin{equation}
M_{\chi^0} = 
\left(\begin{array}{cccc}
M_1 & 0 & -\frac{g^\prime}{\sqrt{2}}v_d & \frac{g^\prime}{\sqrt{2}}v_u \\
0 & M_2 & \frac{g}{\sqrt{2}}v_d & -\frac{g}{\sqrt{2}}v_u \\
-\frac{g^\prime}{\sqrt{2}}v_d & \frac{g}{\sqrt{2}}v_d & 0 & -\mu \\
\frac{g^\prime}{\sqrt{2}}v_u & -\frac{g}{\sqrt{2}}v_u & -\mu & 0
\end{array}\right)_{4\times 4},
\end{equation}
and
\begin{eqnarray}
M_{N-N} = (2\kappa^{ijk}n_i)_{3\times 3}.
\end{eqnarray}
Here $M_1,M_2$ are the U(1) and SU(2) 
 gaugino masses respectively. Now the 
Dirac-type masses are given by the $3\times 7$ matrix.
\begin{equation}
m^T = 
\left(\begin{array}{ccc}
0 & 0 & 0 \\
0 & 0 & 0 \\
0 & 0 & 0 \\
 & Y^{ij}_\nu n_j - \epsilon_i & \\
 & Y^{ij}_\nu v_u 
\end{array}\right)_{7\times 3}.
\end{equation}

Thus the seesaw part of the light neutrino matrix is completely specified by
our superpotential and the scalar vev's, which in turn are derived from
the effective Lagrangian in the intermediate scale of SUSY breaking.

\subsection{Radiative masses}
Given the low-energy Lagrangian emerging in our theory, the diagrams that can
contribute to neutrino masses are given in figure 1. Out of them, the 
dominant contribution comes from 1(a), yielding neutrino mass of the form
\begin{figure}
\begin{center}
\includegraphics[]{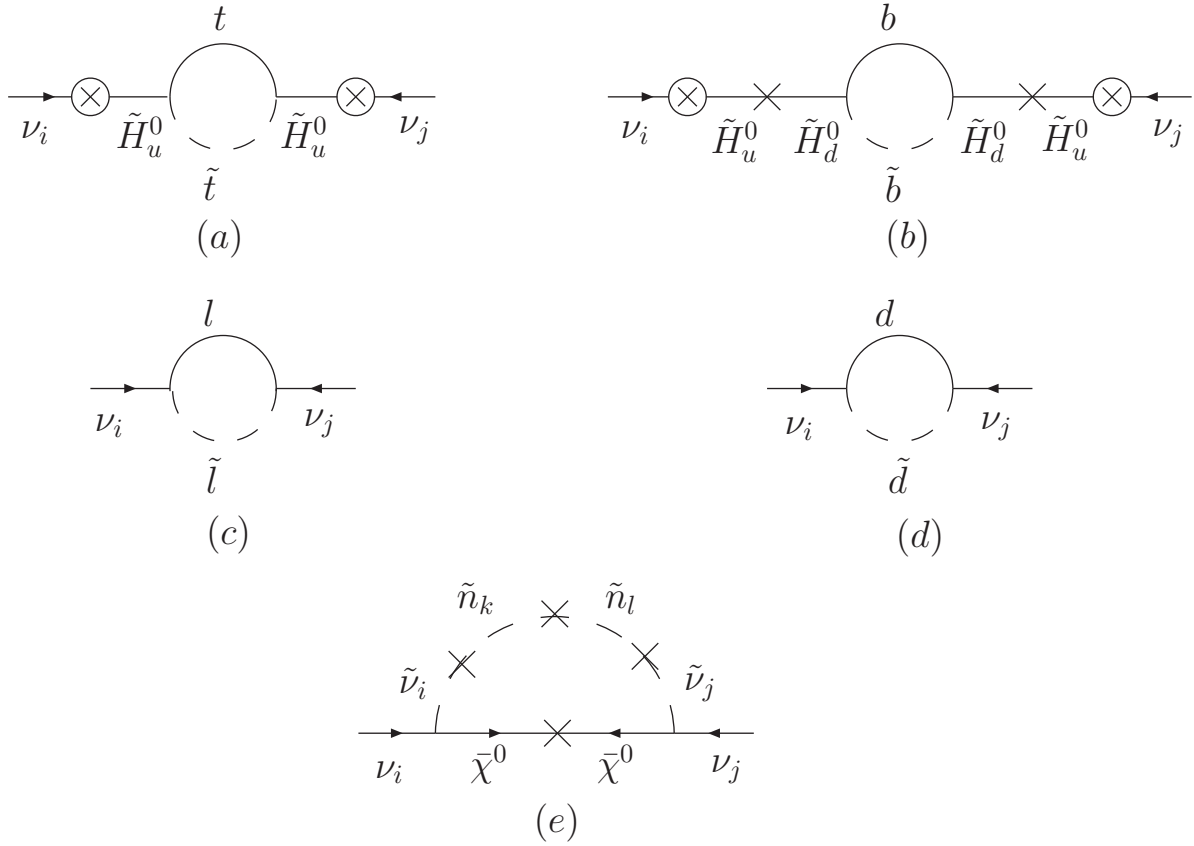}
%\vskip-8cm
\caption{One loop diagrams for neutrino masses. Loops in (a), (b), (c), and 
(d) are induced by fermion and its superpartner. In (e) it is induced by 
neutralino, left- and right- chiral sneutrinos.}
\end{center}
\end{figure}
\begin{equation}
(m_{\nu}^r)_{ij} = \frac{3y_t^2}{32\pi^2}\frac{m_t\sin 2\phi_t}{(m_{H_u^0})^2}
\left\{\frac{M_{P_2}^2}{m_t^2-M_{P_2}^2}\ln \frac{m_t^2}{M_{P_2}^2} - 
\frac{M_{P_1}^2}{m_t^2-M_{P_1}^2}\ln \frac{m_t^2}{M_{P_1}^2}\right\}
\epsilon_i^\prime \epsilon_j^\prime ,
\end{equation}
where $\epsilon^\prime_i=Y_\nu^{ij}n_j-\epsilon_i$, $y_t$ is top 
Yukawa coupling, $\phi_t$ is the angle of 
left-right mixing of stop states and $M_{P_1},M_{P_2}$ are two mass 
eigenvalues of stops. Diagram 1(b) requires two more mass insertions plus the
replacement of $y_t(m_t)$ by $y_b(m_b)$, causing a relative suppression. 
1(c) and 1(d) also give smaller contributions \cite{gh}, since they are 
bilinear in
$\lambda_{ijk}$ and $\lambda^\prime_{ijk}$, which are restricted to rather
small values due to the constraint on left-sneutrino vev's \cite{rm}. 
And finally, 
diagram 1(e) depends on the value of the soft parameters $A$. However, while
in reference \cite{mrs} it could be as big as on the TeV scale, here the 
parameter 
is of much smaller magnitude. This is because the major contribution
from this diagram in the earlier case came to be proportional to
$A_\nu^2$, while $A_\nu \sim F_X/M_P = 0$ in this case. The remaining
contributions, which arise essentially from the F-terms 
and are proportional to the neutrino Yukawa couplings,
are found to be very small.
Thus the contribution from diagram 1(e), too, is of a subleading nature,
and the radiative contribution to neutrino mass matrix is faithfully
given by equation (25).

\subsection{Analysis using the observed pattern}

Our purpose is to show that the scenario proposed above has a solution
space with parameters which are consistent with the proposal that all new 
physics in the observable sector is in the Tev scale. For this,
all we do is to show a few restricted cases where consistent solutions
are found.

We make the following simplification to get a diagonal mass
matix for the right-handed neutrinos:
\begin{equation}
\kappa^{111}=\kappa^{222}=\kappa^{333}=\frac{1}{6}
\end{equation}
and all other $\kappa^{ijk}=0$. We can simplify our analysis further
(without losing the general nature of the conclusions) by assuming
the same vev for all three right-chiral sneutrinos, namely,  
$n_1=n_2=n_3=n$. With this simplification, the total neutrino 
mass matrix is given by 
\begin{eqnarray}
m_\nu &=& m_\nu^s + m_\nu^r
	\nonumber \\
	&=& a \left(\begin{array}{ccc}
\epsilon_1^\prime\epsilon_1^\prime & \epsilon_1^\prime\epsilon_2^\prime & 
\epsilon_1^\prime\epsilon_3^\prime \\
\epsilon_2^\prime\epsilon_1^\prime & \epsilon_2^\prime\epsilon_2^\prime & 
\epsilon_2^\prime\epsilon_3^\prime \\
\epsilon_3^\prime\epsilon_1^\prime & \epsilon_3^\prime\epsilon_2^\prime & 
\epsilon_3^\prime\epsilon_3^\prime \end{array}\right)
-b \left(\begin{array}{ccc}
Y_\nu^{11} & Y_\nu^{12} & Y_\nu^{13} \\
Y_\nu^{12} & Y_\nu^{22} & Y_\nu^{23} \\
Y_\nu^{13} & Y_\nu^{23} & Y_\nu^{33}
\end{array}\right)\cdot
\left(\begin{array}{ccc}
Y_\nu^{11} & Y_\nu^{12} & Y_\nu^{13} \\
Y_\nu^{12} & Y_\nu^{22} & Y_\nu^{23} \\
Y_\nu^{13} & Y_\nu^{23} & Y_\nu^{33}
\end{array}\right)
\end{eqnarray}
\begin{eqnarray}
a &=& \frac{(g^2M_1+g^{\prime 2}M_2)v_d^2}{2(-\mu^2M_1M_2+g^2\mu M_1v_uv_d+
g^{\prime 2}\mu M_2v_uv_d)} + 
	\nonumber \\
&&\frac{3y_t^2}{32\pi^2}\frac{m_t\sin 2\phi_t}
{(m_{H_u^0})^2}\left\{\frac{M_{P_2}^2}{m_t^2-M_{P_2}^2}
\ln\frac{m_t^2}{M_{P_2}^2} - \frac{M_{P_1}^2}{m_t^2-M_{P_1}^2}
\ln\frac{m_t^2}{M_{P_1}^2}\right\}, 
	\nonumber \\
b &=& \frac{v_u^2}{n}.
\end{eqnarray}
A focal theme of our discussion is that mixing in the neutrino sector
is quite different from the quark sector; thus while the latter have
(to the lowest approximation) near-diagonal Yukawa couplings, the former 
can have ample non-diagonality there. In order to accentuate this point,
we may consider the extreme case where the $Y^{ij}_\nu$'s are non-vanishing
only for $i \neq j$. This reduces the number of unknown variables in
our equations, and makes it simpler to illustrate our points. It should,
however, be remembered that this corresponds to a subset of the allowed
parameter space. If the digonal $Y^{ij}_\nu$'s are also admitted, then 
a larger volume in this space will be allowed, demonstrating an even better
viability of the scenario.

After substituting equation (16) into (27), we obtain the six equations 
given below:
\begin{eqnarray}
\frac{2}{3}m_1+\frac{1}{3}m_2 &=& a\epsilon_1^\prime\epsilon_1^\prime 
-b((Y_\nu^{12})^2+(Y_\nu^{13})^2),
	 \\
\frac{1}{6}(m_1+2m_2+3m_3) &=& a\epsilon_2^\prime\epsilon_2^\prime 
-b((Y_\nu^{12})^2+(Y_\nu^{23})^2),
	 \\
&=& a\epsilon_3^\prime\epsilon_3^\prime - b((Y_\nu^{13})^2+(Y_\nu^{23})^2),
	 \\
\frac{1}{3}(-m_1+m_2) &=& a\epsilon_1^\prime\epsilon_2^\prime -
bY_\nu^{13}Y_\nu^{23},
	 \\
\frac{1}{3}(m_1-m_2) &=& a\epsilon_1^\prime\epsilon_3^\prime -
bY_\nu^{12}Y_\nu^{23},
	 \\
\frac{1}{6}(-m_1-2m_2+3m_3) &=& a\epsilon_2^\prime\epsilon_3^\prime -
bY_\nu^{12}Y_\nu^{13}.
\end{eqnarray}
For fixed mass eigenvalues, the above equations can 
be solved for the six parameters: $\epsilon_1^\prime ,\epsilon_2^\prime ,
\epsilon_3^\prime, Y_\nu^{12},Y_\nu^{13}, Y_\nu^{23}$. 
Using the solar neutrino data, one can use 
$m_1 \simeq m_2$ in the lowest approximation. Thus, at zeroth order,
we set the left-hand sides in equations (32) and (33) equal to zero. In that
case, it is possible to choose the solutions 
$\epsilon_2^\prime = \epsilon_3^\prime$ and $Y_\nu^{12} = Y_\nu^{13}$ 
(which is consistent with maximal 23-mixing), thereby reducing the
six equations to four only. Of course, we get confined to an even smaller
part of the entire allowed paramters space. But our conclusions
still remain quite general, there being small corrections to the solutions 
when the non-zero value of $\Delta m^2_{12}$ is inserted.

The equations in the simplified form are 

\begin{eqnarray}
m_1 = m_2 = -b(Y_\nu^{23})^2,
	\nonumber \\
m_3 = a(2(\epsilon_2^\prime )^2-(\epsilon_1^\prime )^2) - 2|m_2|,
	\nonumber \\
a\epsilon_1^\prime \epsilon_2^\prime = Y_\nu^{12}\sqrt{b|m_2|},
	\nonumber \\
\frac{a^2}{|m_2|}(\epsilon_1^\prime )^4 + a\left[1+\frac{m_3}{|m_2|}\right]
(\epsilon_1^\prime )^2 - |m_2| = 0.
\end{eqnarray}
where one should note that each of the 
first and second mass eigenvalues has Majorana phase equal to $\pi$. 

From the four equations given above, we can determine 
the four parameters still remaining independent: 
$\epsilon_1^\prime ,\epsilon_2^\prime ,
Y_\nu^{12},Y_\nu^{23}$. We simultaneously check that the EWSB conditions 
which are listed in section 2.2 are satisfied in parameter space 
we are led into. Some sample solutions thus obtained have been presented 
in the next section.

It should however, be kept in mind that we have made use of just a few
hidden sector superfields in the effective low-energy theory. Therefore,
the treatment described here cannot address potentially destabilising
higher order effects in the final supergravity framework. Thus, the
relationships among the numerical values of various parameters should be
treated as indicative ones only.

We also admit that minimization of only the tree-level potential has been
considered here, and the conditions are obviously going to change when
loop corrections are taken into account. However, even then the general
conclusion that neutrino masses and mixing can be governed by TeV scale
physics, mediated via right-chiral neutrino superfields and electroweak
symmetry breaking conditions, continues to  be true in this sample
scenario allowing $\Delta L = 1$.

\section{Numerical results}

To solve for the four unknown parameters in equation (35), we need to know the 
mass eigenvalues of neutrinos. So far, we do not know the 
exact values of neutrino masses. From the available data, which suggest
neutrino oscillations, the following mass-squared differences 
are favoured \cite{sv}:
\begin{eqnarray}
\Delta m_{21}^2 &=& (8.0\pm 0.3)\times 10^{-5} {\rm eV}^2,
	\nonumber \\
|\Delta m_{32}^2| &=& (2.5\pm 0.3)\times 10^{-3} {\rm eV}^2.
\end{eqnarray}
The two mass-squared differences shown above indicates three possibilities 
\cite{hier}, namely
\begin{enumerate}
\item
Normal hierarchy: $m_1\approx m_2\sim \sqrt{\Delta m_{21}^2}$, 
$m_3\sim \sqrt{|\Delta m_{32}^2|}$.
\item
Inverted hierarchy: $m_1\approx m_2\sim \sqrt{|\Delta m_{32}^2|}$, 
$m_3\ll \sqrt{|\Delta m_{32}^2|}$.
\item
Degenerate masses: $m_1\approx m_2\approx m_3\gg\sqrt{|\Delta m_{32}^2|}$.
\end{enumerate}
We solve the four equations in equation (35) in each of the above three 
cases for 
the parameters $\epsilon_1,\epsilon_2,Y_\nu^{12},Y_\nu^{23}$. It is 
to be noticed that earlier we chose $\epsilon_2^\prime = \epsilon_3^\prime$ 
and $Y_\nu^{12} = Y_\nu^{13}$. This implies $\epsilon_2 = \epsilon_3$,
a postulate frequently made in bilinear R-parity violation in the light
of maximal $\nu_\mu - \nu_\tau$ mixing. 
We plug these parameters in equation (12) to solve SUSY soft breaking 
parameters. 
Next we check in each case whether the combination of parameters 
ensure that EWSB is triggered by following the prescriptions of section 2.2. 
For simplicity, we have chosen the various SUSY soft breaking parameters to
be of the form:
\begin{eqnarray}
(m_{\tilde{L}}^2)^{ij} = m_{\tilde{L}}^2\delta^{ij}, \quad 
(m_{\tilde{N}}^2)^{ij} = m_{\tilde{N}}^2\delta^{ij}, \quad 
(A_\nu Y_\nu )^{ij} = (A_\nu Y_\nu )\delta^{ij}, 
	\nonumber \\
(A_\kappa\kappa )^{ijk} = 0~{\rm if~any~of~the~indices~are~different~from~
each~other}.
\end{eqnarray}
These forms also ensure the suppression of FCNC. 
In this case one has $(A_\kappa\kappa )^{222} = 
(A_\kappa\kappa )^{333}$ and $(B_\epsilon\epsilon)^2 = 
(B_\epsilon\epsilon)^3$. The numerical values of various standard model 
and MSSM parameters chosen for our analysis are 
\begin{eqnarray}
M_2 = 300~{\rm GeV}, M_1 = 0.5 M_2, m_t = 172.9~{\rm GeV}, 
m_{H_u}^2 = -(300~{\rm GeV})^2, m_{\tilde{L}}^2 = (500~{\rm GeV})^2, 
	\nonumber \\
(A_\nu Y_\nu ) = 10^{-4}~{\rm GeV}, M_{P_2} = 600~{\rm GeV}, 
M_{P_1} = 500~{\rm GeV}, \sin 2\phi_t = 1.
\end{eqnarray}
We present our numerical values in tabular form in each case.

\subsection{Normal hierarchy}

We have taken $m_1\approx m_2 = \sqrt{0.8}\times 10^{-11}~{\rm GeV}$, 
$m_3 = \sqrt{25.}\times 10^{-11}~{\rm GeV}$.  Table 2 contains some 
sample solutions in this scenario.
\begin{table}
\begin{center}
\begin{tabular}{||c||c|c|c|c|c|c|c|c||} \hline \hline

$\tan\beta$ & 8.5 & 8.5 & 8.5 & 8.5 & 8.5 & 8.5 & 8.5 & 8.5 \\ \cline{1-1}
$m_{\tilde{N}}$ 
 & 500.0 & 500.0 & 200.0 & 200.0 & 
500.0 & 500.0 & 200.0 & 500.0 \\ \cline{1-1}
$\mu$ 
 & 300.0 & -300.0 & 300.0 & -300.0 & 500.0 & -500.0 & 500.0 & -500.0 
\\ \cline{1-1}
$n$ 
 & 1000.0 & 1000.0 & 400.0 & 400.0 & 1000.0 & 1000.0 & 400.0 & 400.0 \\ 
\hline
$\epsilon_1$ 
 & -0.0015 & -0.0009 & -0.0021 & -0.0015 & -0.0028 & -0.0022 & 
-0.0034 & -0.0028 \\ \cline{1-1}
$\epsilon_2$ 
 & -0.011 & -0.0079 & -0.0117 & -0.0086 & -0.0175 & -0.0145 & 
-0.0182 & -0.0152 \\ \cline{1-1}
$Y_\nu^{12}$ & & & & & & & & \\
$(\times 10^{-7})$ & 4.147 & 4.147 & 2.623 & 2.623 & 4.147 & 4.147 & 
2.623 & 2.623 \\ \cline{1-1}
$Y_\nu^{23}$ & & & & & & & & \\
$(\times 10^{-7})$ & 5.473 & 5.473 & 3.461 & 3.461 & 5.473 & 5.473 & 
3.461 & 3.461 \\ \hline
$(B_\mu\mu)$ 
 & 34388.7 & 34388.7 & 34388.7 & 34388.7 & 1394388.8 & 
1394388.8 & 1394388.8 & 1394388.8 \\ \cline{1-1}
$m_{H_d}$ 
 & 454.3 & 454.3 & 454.3 & 454.3 & 3406.8 & 3406.8 & 
3406.8 & 3406.8 \\ \cline{1-1}
$(A_\kappa\kappa)^{111}$ 
 & -750.0 & -750.0 & -300.0 & -300.0 & -750.0 & -750.0 & -300.0 & 
-300.0 \\ \cline{1-1}
$(A_\kappa\kappa)^{222}$ 
 & -750.0 & -750.0 & -300.0 & -300.0 & -750.0 & -750.0 & -300.0 & 
-300.0 \\ \cline{1-1}
$(B_\epsilon\epsilon )^1$ 
 & 0.432 & 0.576 & -0.0008 & 0.143 & 0.302 & 0.693 & 
-0.131 & 0.26 \\ \cline{1-1}
$(B_\epsilon\epsilon )^2$ 
 & 0.159 & 0.894 & -0.334 & 0.402 & -0.504 & 1.491 & 
-0.996 & 0.999 \\ \hline \hline

\end{tabular}
\end{center}
\caption{Various sample solutions corresponding to normal hierarchy. All
masses are in GeV.}
\end{table}
In the table, the first four parameters such as $\tan\beta$, 
$m_{\tilde{N}}$, $\mu$ and $n$ are fixed in such a way as to get real 
solutions of equation (35) and also satisfy EWSB conditions. In order to 
have real 
solutions for $\epsilon_1$, $\epsilon_2$, $Y_\nu^{12}$ and $Y_\nu^{23}$ from 
equation (35), we have found that $\tan\beta$ should be at least 8.5, 
independent of $\mu$ and $n$. We have allowed $\mu$ to vary 
from $\pm$300 GeV to $\pm$500 GeV. Among the EWSB conditions, the requirement
that the potential 
at the minima is negative puts the severest constraint on the right-handed 
sneutrino vev,$n$, for a particular value of $m_{\tilde{N}}$. From 
table 2 it can be noticed that for $m_{\tilde{N}}$ of 500, 
the minimum value of $n$ should be around 1000 GeV. If we decrease 
$m_{\tilde{N}}$ to 200, the value of $n$ should be at least 
400 GeV.

This vindicates our earlier statement, based largely on the requirement
of negativity of the scalar potential at the minimum, that $\langle\tilde
{n}_i\rangle$ around the electroweak scale is viable. Even if small values
of $\langle\tilde{n}_i\rangle$ can be engineered via fine tuning, it would
bring light sterile neutrinos into the picture, taking us outside the
ambit of three-flavour tri-bimaximal mixing. Such a digression is avoided
in this study.

For each case one can also evaluate the various soft SUSY 
breaking parameters like $(B_\mu\mu)$, $m_{H_d}$, 
$(A_\kappa\kappa)^{111}$, $(A_\kappa\kappa)^{222}$, $(B_\epsilon\epsilon)^1$, 
$(B_\epsilon\epsilon)^2$, using equation (12). One should perhaps note that
the $\epsilon$-parameters are of rather small values due to the restrictions
on left-chiral sneutrino vev's, a constraint arising from neutrino masses
and widely used in works on R-parity violating SUSY. If we factor out such
small values, the SUSY-breaking parameters including $B_\epsilon$ are
roughly around the TeV range, as expected.

\subsection{Inverted hierarchy}

In this case we have taken 
$m_1\approx m_2 = \sqrt{25.}\times 10^{-11}~{\rm GeV}$,
$m_3 = 10^{-3}\times m_2$. Some sample solutions in the
parameter space are presented in table 3. 
\begin{table}
\begin{center}
\begin{tabular}{||c||c|c|c|c|c|c|c|c||} \hline \hline

$\tan\beta$ & 8.5 & 8.5 & 8.5 & 8.5 & 8.5 & 8.5 & 8.5 & 8.5 \\ \cline{1-1}
$m_{\tilde{N}}$ 
 & 500.0 & 500.0 & 200.0 & 200.0 &
500.0 & 500.0 & 200.0 & 200.0 \\ \cline{1-1}
$\mu$ 
 & 300.0 & -300.0 & 300.0 & -300.0 & 500.0 & -500.0 & 500.0 & -500.0 
\\ \cline{1-1}
$n$ 
 & 1000.0 & 1000.0 & 400.0 & 400.0 & 1000.0 & 1000.0 & 400.0 & 400.0 \\
\hline
$\epsilon_1$ 
 & -0.009 & -0.006 & -0.011 & -0.0078 & -0.015 & -0.012 &
-0.017 & -0.014 \\ \cline{1-1}
$\epsilon_2$ 
 & -0.014 & -0.0097 & -0.0158 & -0.0116 & -0.0229 & -0.0188 &
-0.0247 & -0.0207 \\ \cline{1-1}
$Y_\nu^{12}$ & & & & & & & & \\
$(\times 10^{-6})$ & 1.164 & 1.164 & 7.36 & 7.36 & 1.164 & 1.164 &
7.36 & 7.36 \\ \cline{1-1}
$Y_\nu^{23}$ & & & & & & & & \\
$(\times 10^{-6})$ & 1.294 & 1.294 & 8.184 & 8.184 & 1.294 & 1.294 &
8.184 & 8.184 \\ \hline
$(B_\mu\mu)$ 
 & 34388.7 & 34388.7 & 34388.7 & 34388.7 & 1394388.8 &
1394388.8 & 1394388.8 & 1394388.8 \\ \cline{1-1}
$m_{H_d}$ 
 & 454.3 & 454.3 & 454.3 & 454.3 & 3406.8 & 3406.8 &
3406.8 & 3406.8 \\ \cline{1-1}
$(A_\kappa\kappa)^{111}$ 
 & -750.0 & -750.0 & -300.0 & -300.0 & -750.0 & -750.0 & -300.0 &
-300.0 \\ \cline{1-1}
$(A_\kappa\kappa)^{222}$ 
 & -750.0 & -750.0 & -300.0 & -300.0 & -750.0 & -750.0 & -300.0 &
-300.0 \\ \cline{1-1}
$(B_\epsilon\epsilon )^1$ 
 & 0.865 & 1.6 & -0.241 & 0.454 & 0.239 & 2.124 &
-0.867 & 1.018 \\ \cline{1-1}
$(B_\epsilon\epsilon )^2$ 
 & 0.748 & 1.76 & -0.417 & 0.595 & -0.164 & 2.581 &
-1.328 & 1.417 \\ \hline \hline

\end{tabular}
\end{center}
\caption{Various sample solutions corresponding to inverted hierarchy. All
masses are in GeV.}
\end{table}
As in the case of normal hierarchy, here also we have found a minimum required 
value of 8.5 for $\tan\beta$ in order to get real solutions of 
equation (35). The condition that the potential at the minima should 
be less than zero puts lower limit on $n$ for a particular value 
of $m_{\tilde{N}}$. The least values of $n$ thus obtained have
been presented in solutions, here as well as in the tables 2 and 4.
Except the parameters $(B_\mu\mu)$, $m_{H_d}$, $(A_\kappa\kappa)^{111}$ 
and $(A_\kappa\kappa)^{222}$, all other parameters 
in inverted hierarchy are  different from the previous case 
because of different neutrino mass eigenvalues.

\subsection{Degenerate neutrinos}

For this case we have used 
$m_1\approx m_2\approx m_3 = 10^{-10}~{\rm GeV}$.
Table 4 contains some sample solutions.
\begin{table}
\begin{center}
\begin{tabular}{||c||c|c|c|c|c|c|c|c||} \hline \hline

$\tan\beta$ & 5 & 5 & 5 & 5 & 1.5 & 2.5 & 1.5 & 2.5 \\ \cline{1-1}
$m_{\tilde{N}}$ 
 & 500.0 & 500.0 & 200.0 & 200.0 &
500.0 & 500.0 & 200.0 & 200.0 \\ \cline{1-1}
$\mu$ 
 & 300.0 & -300.0 & 300.0 & -300.0 & 500.0 & -500.0 & 500.0 & -500.0 
\\ \cline{1-1}
$n$ 
 & 1000.0 & 1000.0 & 400.0 & 400.0 & 1000.0 & 1000.0 & 400.0 & 400.0 \\
\hline
$\epsilon_1$ 
 & 0.0006 & 0.0001 & -0.0018 & -0.0022 & 0.0013 & -0.0094 &
-0.0015 & -0.0118 \\ \cline{1-1}
$\epsilon_2$ 
 & -0.0018 & -0.0026 & -0.0043 & -0.0052 & -0.0008 & -0.022 &
-0.0038 & -0.0248 \\ \cline{1-1}
$Y_\nu^{12}$ & & & & & & & & \\
$(\times 10^{-6})$ & 1.559 & 1.559 & 0.986 & 0.986 & 1.837 & 1.646 &
1.162 & 1.041 \\ \cline{1-1}
$Y_\nu^{23}$ & & & & & & & & \\
$(\times 10^{-6})$ & 1.853 & 1.853 & 1.172 & 1.172 & 2.184 & 1.957 &
1.381 & 1.238 \\ \hline
$(B_\mu\mu)$ 
 & 19196.7 & 19196.7 & 19196.7 & 19196.7 & 242399.6 &
407529.8 & 242399.6 & 407529.8 \\ \cline{1-1}
$m_{H_d}$ 
 & 99.1 & 99.1 & 99.1 & 99.1 & 339.4 &
878.5 & 339.4 & 878.5 \\ \cline{1-1}
$(A_\kappa\kappa)^{111}$ 
 & -750.0 & -750.0 & -300.0 & -300.0 & -750.0 & -750.0 & -300.0 &
-300.0 \\ \cline{1-1}
$(A_\kappa\kappa)^{222}$ 
 & -750.0 & -750.0 & -300.0 & -300.0 & -750.0 & -750.0 & -300.0 &
-300.0 \\ \cline{1-1}
$(B_\epsilon\epsilon )^1$ 
 & 1.505 & 1.837 & 0.045 & 0.376 & 1.141 & 4.274 &
-0.57 & 2.735 \\ \cline{1-1}
$(B_\epsilon\epsilon )^2$ 
 & 1.495 & 2.168 & -0.098 & 0.574 & 0.494 & 7.034 &
-1.373 & 5.355 \\ \hline \hline

\end{tabular}
\end{center}
\caption{Various sample solutions corresponding to degenerate neutrinos. All
masses are in GeV.}
\end{table}
Unlike the two previous cases, we have got different $\tan\beta$ values 
here. We have found that for $\mu=\pm$300 GeV, $\tan\beta$ 
should be at least 5 in order that equation (35) yields real solutions for 
$\epsilon_i$ and $Y^{ij}_\nu$. Moreover, for a $\mu~=$  500 GeV and 
-500 GeV, we have found that the minimum of $\tan\beta$  should be around 
1.5 and 2.5 respectively, the former of which is practicaly ruled out
from the Large Electron Positron (LEP) collider data. 
We find that for large $\mu$, the 
minimum allowed value of $\tan\beta$ is sensitive to the sign of $\mu$. 
This kind of 
feature is not there in the two previous cases. The EWSB conditions give 
similar constraints as in the previous two cases.

\section{Summary and conclusions}

In this work we have studied R-parity violating effects on neutrino masses. 
The special features of neutrinos are postulated to arise from 
nonrenormalizable interactions with hidden sector fields. 
Proposing right-chiral 
Majorana-like neutrino superfields and choosing specific R-charges of 
hidden and visible sector superfields, we could construct a term in the 
superpotential, which violates lepton number by three units and thus violates 
R-parity explicitly. 

Comparing our model with reference \cite{mrs}, an interesting dichotomy can be
observed. Whereas non-vanishing of $F_X$-values have a crucial role in the
previous case, they are forced to be zero here. The role has now shifted to
the $\Delta L = 3$ term in the superpotential, which is allowed by our
R-charge assignments.

It has been further argued by us that lepton
number violation, distinct R-charges for the right-chiral (s)neutrinos and
special properties of the hidden sector fields they couple to --- 
all have
led together to the distinctive features of neutrino masses and mixing.
After SUSY breaking and EWSB, neutrino masses of the right order are generated 
from seesaw and radiative effects, provided that the vev's of 
right-chiral sneutrinos are of order 1 TeV. The model is also subjected to 
conditions arisng from EWSB. Finally, we succeed in demonstrating that
viable solutions exist, with the requisite mixing pattern, for each of 
the cases of normal hierarchy, inverted hierarchy and degenerate neutrinos.
The scheme to link features of the neutrino sector to the TeV scale,
initiated in a previous work by us, is thus completed.

\end{document}